\documentclass[12pt,preprint,apj]{emulateapj}

\newcommand{\mbh}{M_{\rm BH}}
\newcommand{\msigma}{M$_{\rm BH}$-$\sigma_*$}
\newcommand{\msun}{M_{\rm \sun}}
\newcommand{\sigstar}{\sigma_{*}}

\newcommand{\reff}{R_{\rm e}}
\newcommand{\kms}{km s$^{-1}$}

\shorttitle{}
\shortauthors{Woo et al.}

\begin{document}

\title{DO QUIESCENT AND ACTIVE GALAXIES HAVE DIFFERENT M$_{BH}-\sigma_*$ RELATIONS?}

\author{JONG-HAK WOO$^{1,2}$\altaffilmark{,7}}
\author{ANDREAS SCHULZE$^{3}$}
\author{DAESEONG PARK$^{1}$}
\author{WOL-RANG KANG$^{1}$}
\author{SANG CHUL KIM$^{4}$}
\author{DOMINIK A. RIECHERS$^{5,6}$}

\affil{$^{1}$Astronomy Program, Department of Physics and Astronomy, Seoul National University, 1 Gwanak-ro Gwanak-gu, Seoul, 151-742, Republic of Korea; woo@astro.snu.ac.kr}
\affil{$^{2}$The Observatories of the Carnegie Institution for Science, 813 Santa Barbara St. Pasadena, CA 91101, USA}
\affil{$^{3}$Kavli Institute for Astronomy and Astrophysics, Peking University, 100871 Beijing, China}
\affil{$^{4}$Korea Astronomy and Space Science Institute, Daejeon 305-348, Republic of Korea}
\affil{$^{5}$Astronomy Department, Cornell University, 220 Space Science Building, Ithaca, NY 14853, USA}
\affil{$^{6}$Astronomy Department, California Institute of Technology, MC 249-17, 1200 East California Boulevard, Pasadena, CA 91125, USA}
\altaffiltext{7}{TJ Park Science Fellow}

\begin{abstract}
To investigate the validity of the assumption that quiescent galaxies and active galaxies 
follow the same black hole mass ($\mbh$)- stellar velocity dispersion ($\sigstar$) relation,
as required for the calibration of $\mbh$ estimators for broad line AGNs,
we determine and compare the \msigma\ relations, respectively, for quiescent and 
active galaxies. For the quiescent galaxy sample, composed of 72 dynamical $\mbh$ measurements,
we update $\sigstar$ for 28 galaxies using homogeneous H-band measurements 
that are corrected for galaxy rotation. For active galaxies, we collect 25 reverberation-mapped
AGNs and improve $\sigstar$ measurement for two objects . 
Combining the two samples, we determine the virial factor $f$, 
first by scaling the active galaxy sample to the \msigma\ relation of quiescent galaxies, and second 
by simultaneously fitting the quiescent and active galaxy samples,
as $f=5.1_{-1.1}^{+1.5}$ and $f=5.9_{-1.5}^{+2.1}$, respectively. 
The \msigma\ relation of active galaxies appears to be shallower than that of quiescent galaxies.
However, the discrepancy is caused by a difference in the accessible $\mbh$ distribution at 
given $\sigstar$, primarily due to the difficulty of measuring reliable stellar velocity dispersion 
for the host galaxies of luminous AGNs. Accounting for the selection effects, we find that 
active and quiescent galaxies are consistent with following intrinsically the same \msigma\ relation.

\end{abstract}



\section{Introduction}
Supermassive black holes (BHs) and their host galaxies appear to coevolve as manifested 
by the present-day correlations between BH mass ($\mbh$) and galaxy properties, i.e., $\mbh$ - stellar velocity
dispersion ($\sigstar$) relation \citep{Ferrarese:2000,Gebhardt:2000a,Gultekin:2009,McConnell:2013}.
These relations are established based on $\mbh$ measurements from 
spatially resolved kinematics of stars \citep[e.g.][]{vanderMarel:1998,Gebhardt:2000a}, 
gas \citep[e.g.,][]{Ferrarese:1996,Marconi:2001} or 
maser \citep{Herrnstein:2005,Kuo:2011} around BH's sphere of influence. 
While in the early studies, the BH-galaxy relations appeared to be very tight, 
even consistent with zero intrinsic scatter \citep{Gebhardt:2000a}, the increased dynamical
range and sample size revealed a larger intrinsic scatter of $\sim0.5$~dex \citep{Gultekin:2009}
and significant outliers from the relation \citep[e.g.,][]{Hu:2008,Greene:2010,Kormendy:2011,vandenbosh:2012}. 

Dynamical $\mbh$ measurements based on the spatially resolved kinematics are 
challenging for broad line Active Galactic Nuclei (AGNs), 
due to the presence of a bright nuclear point source. Instead, an alternative method to measure $\mbh$, 
namely reverberation mapping has been adopted for broad line AGNs \citep[e.g.][]{Peterson:1993}. 
By employing the variability of AGN, this method measures the time delay between continuum and broad line variability, corresponding to the broad line region (BLR) size ($R_{\rm BLR}$). The black hole mass is then derived via the virial relation $\mbh=f \Delta V^2 R_{\rm BLR} / G$, where G is the gravitational constant. 
The broad line cloud velocity $\Delta V$ is  measured from the broad line width, and $f$ is a scale factor that depends on the geometry and kinematics of the BLR 
gas, which is poorly known for individual objects (cf. Brewer et al. 2011; Pancost et al. 2012). 
Thus, broad line AGNs are the only available probes to determine $\mbh$ out to
cosmological distances and test for evolution of the BH-galaxy correlations \citep[e.g.][]{Peng:2006,Woo:2006,Woo:2008,Merloni:2010,Decarli:2010,Cisternas:2011}.

Before using AGNs to measure possible redshift evolution in the \msigma\ relationship, it is
necessary to verify whether AGNs follow the same \msigma\ relationship as quiescent galaxies in the local universe. To first order this seems to be the case, as indicated by the early studies of small samples \citep{Gebhardt:2000b,Ferrarese:2001,Nelson:2004}. Motivated by these results, \citet{Onken:2004} \emph{assumed} that broad line AGNs obey the same \msigma\ relationship as quiescent galaxies and then empirically determine the unknown factor $f=5.5\pm1.8$, by normalizing the reverberation AGNs to the \msigma\ relation of quiescent galaxies. 
Using a significantly increased sample size and dynamical range, \citet{Woo:2010} redetermined the scale 
factor $f=5.2\pm1.2$, and directly fitted the \msigma\ relation for reverberation-mapped AGN, finding a slope $\beta=3.55\pm0.6$ in the \msigma\ relation, 
\begin{equation}
\log (\mbh / \msun) = \alpha + \beta \log (\sigma_{\ast} / 200\ \mathrm{km~s^{-1}}), 
\end{equation}
which was shallower than the slope $\beta=4.24\pm0.41$ for quiescent galaxies reported by \citet{Gultekin:2009}, 
but consistent within the uncertainties. By adding new $\mbh$ measurements of most massive BHs, 
\citet{McConnell:2013} recently presented an updated \msigma\ 
relation for quiescent galaxies, with a significantly steeper slope of $\beta=5.64\pm0.32$. 

Therefore, the slopes of \msigma\ relation between quiescent and active galaxies are now only consistent 
with each other on the $2\sigma$ level. The observed deviation can be either caused by a pure statistical 
effect or  sample selection, or it may point to a real physical difference of BH - galaxy coevolution 
between active and predominantly quiescent galaxies. Thus, differentiating between these possibilities will 
shed light on the coevolution process. Furthermore, since the assumption of similar \msigma\ relations for 
quiescent and active galaxies is one of the foundations of the virial $\mbh$ estimators,
it is required to reinvestigate and compare the \msigma\ relations of quiescent and active galaxies.

In this paper, we provide updated results on the \msigma\ relations for quiescent and active galaxies. 
The quiescent galaxy sample is mainly based on \citet{McConnell:2013}, but we added improved and homogeneous near-IR $\sigstar$ measurements 
for a large fraction of the sample from \citet[][hereafter K13]{Kang:2013}. The active galaxy sample 
is based on the reverberation-mapped AGNs from \citet{Woo:2010}, with the addition of Mrk~50 and 
updated $\mbh$ and new $\sigstar$ measurements for part of the sample. 
Near-IR spectroscopic observations are presented in Section~\ref{section:obs}, while stellar velocity dispersion measurements are discussed in Section~\ref{section:anal}. We present and discuss 
our results on the \msigma\ relations for quiescent and for active galaxies in Section~\ref{section:msig} and  investigate whether quiescent and active galaxies have consistent \msigma\ relations.
Conclusions are presented in Section~\ref{section:conclu}.


\begin{deluxetable*}{lcccrcrcccccc}
\tablecolumns{13}
\tablewidth{0pc}
\tablecaption{Sample and Observations}
\tablehead{
\colhead{Name} & \colhead{Obs. Date} & \colhead{RA} & \colhead{DEC} & \colhead{EXPT} & \colhead{S/N} &   \colhead{Dist.} & \colhead{Spatial Scale} & \colhead{$R_e$} & \colhead{Ref.} & \colhead{$\sigma_{\rm opt}$} & \colhead{Ref.} & \colhead{$\sigma_{\rm IR}$} \\
\colhead{ }&\colhead{ }&\colhead{ }&\colhead{ }&\colhead{(s)}&\colhead{ }&\colhead{(Mpc)}& \colhead{(kpc/1{\arcsec})} & \colhead{(kpc)} &\colhead{ }&\colhead{(km s$^{-1}$)} &\colhead{ } &\colhead{(km s$^{-1}$)}\\
\colhead{(1)}&\colhead{(2)}&\colhead{(3)}&\colhead{(4)}&\colhead{(5)}&\colhead{(6)}&\colhead{(7)}&\colhead{(8)}&\colhead{(9)}&\colhead{(10)}&\colhead{(11)}&\colhead{(12)}&\colhead{(13)}
}
\startdata
Akn 120 &2010-01-01& 05 16 11.48 & $-$00 09 00.6 & 3,600 &428&141.6& 0.68 & - &-& $221{\pm 17}$ &2& $192{\pm 8}$\\
NGC 3227 &2010-01-01& 10 23 30.58 & $+$19 51 54.2 & 600 &230&17.0& 0.08 & 0.27 &1& $136{\pm 4}$&2& $92{\pm 6}$\\
3C 390.3 &2009-05-22& 18 42 08.99 & $+$79 46 17.1 & 3,200 &520&247.3& 1.17 & - &-& $273{\pm 16}$ &2& $267{\pm 14}$
\enddata
\label{tab:sample}
\tablecomments{Col. (1): galaxy name. Col. (2): observation date. Col. (3) Right Ascension (J2000). Col. (4): Declination (J2000). Col. (5): total exposure time. Col. (6): signal-to-noise ratio per pixel in the extracted spectrum. Col. (7): distance. Col. (8): spatial scale. Col. (9)-(10): effective radius and reference. Col. (11)-(12) : the optical stellar velocity dispersion and reference. Col. (13) : the near-IR stellar velocity dispersion.\\
References. --- (1) \citet{Davies:2006}; (2) \citet{Nelson:2004}}
\end{deluxetable*}


\section{Observations and Data Reduction} \label{section:obs}

To uniformly measure and improve $\sigstar$ using near-IR spectroscopy,
we observed 28 quiescent galaxies and 3 reverberation-mapped AGNs. 
Observation, data reduction and analysis are presented in detail in K13. 
Here, we briefly describe observations for 3 AGNs, which have $\mbh$ determined from 
reverberation mapping results, namely NGC~3227, Akn~120 and 3C~390.3. The stellar velocity dispersions of their host galaxies have been previously measured in the optical, using the Ca II triplet line \citep{Nelson:2004,Onken:2004}. For broad line AGNs, $\sigstar$ measurements of the host galaxies become increasingly difficult with increasing nuclear-to-host flux ratio due to dilution, in particular in the optical. Measurements in the near-IR have the advantage of an improved nuclear-to-host contrast. 

The previous $\sigstar$ measurements of Akn~120 and 3C~390.3 were 
based on low signal-to-noise optical spectra, hence, our near-IR observations could potentially provide a significant improvement on the $\sigstar$ measurement. 
The $\sigstar$\ of NGC~3227 was measured using a fairly good optical spectrum, but we provide here spatially resolved spectroscopy and are therefore able to correct for the galaxy's rotation component (see below).

Observations were performed with the Triplespec at the Palomar Hale 5~m telescope \citep{Wilson:2004}. Triplespec is a near-IR spectrograph with simultaneous coverage from 1.0 $\mu$m to 2.4 $\mu$m at a spectral resolution of R$=2500-2700$. We used a long-slit with 1$\arcsec$ $\times$ 30$\arcsec$, placed along the galaxy's major axis. We performed an ABBA dither pattern along the slit to improve the sky subtraction. For each observing night we observed several A0V stars as telluric standards. We also observed stars of different  spectral types as template stars for the $\sigstar$ measurement (see K13 for details).

We carried out standard data reduction tasks using IRAF scripts, including bias subtraction, flat-fielding, wavelength calibration and extraction of one-dimensional spectra. For Akn~120 and 3C~390.3 we extracted  the spectra within the largest possible aperture ($\pm$7$\arcsec$) to increase the signal-to-noise. For NGC~3227 we extracted spectra from 13 small extraction windows of the same size along the galaxy's major axis,
in order to obtain spatially resolved $\sigstar$.
We also extracted spectra from apertures of increasing size, to test for the effect of aperture size on 
our $\sigstar$ measurement.

We corrected the spectra for telluric lines, following the method given in K13. Several A0V stars were used to generate a telluric template for each observing night. Averaging over several stars removes small variations between individual stars. The mean telluric template for each night was than used to subtract the telluric lines from the spectra.


\section{Stellar Velocity Dispersions} \label{section:anal}
We measured $\sigstar$ in the H band, covering several CO lines, Mg~I and Si~I simultaneously, by 
directly fitting the observed spectra in pixel space to stellar templates, which were broadened by a 
Gaussian velocity ranging from 50 to 350 km~s$^{-1}$. The $\chi^2$ minimization 
was performed using the Gauss-Hermite Pixel Fitting software \citep{vanderMarel:1994,Woo:2005}, after
masking out AGN emission (i.e., \ion{Fe}{2} at 1.65 $\micron$), bad pixels and sky subtraction residuals.
We matched the continuum shapes of the broadened templates and the galaxy spectra by fitting low-order polynomials (order 2-4). 

To account for template mismatch, we measured $\sigstar$ using various M-type template stars (M0~III, M1~III, M2~III, M3~III) since these spectral types provide a consistent match to the observed
galaxy spectra in the H-band. We have also used several K giant stars, but 
found a worse agreement with the observed galaxy spectra. Thus, we have excluded them in the further analysis (see K13 for details). 
The final $\sigstar$ measurements are given by the mean of the measured values from each M-type template,
while the quoted uncertainties are determined by adding the standard deviation of the measurements from 
each template to the mean measurement error in quadrature.

\begin{figure}
\centering
\includegraphics[width = 0.48 \textwidth]{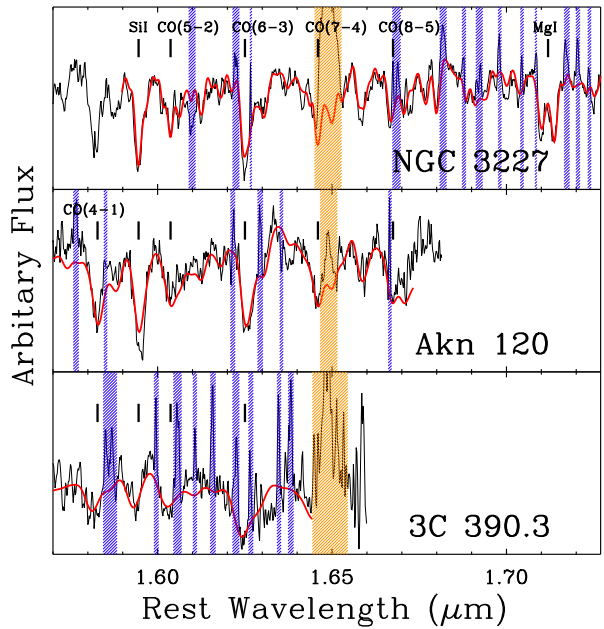}
\caption{Normalized H-band spectra of 3 active galaxies (black line) overplotted with the 
best-fit model (red line). Sky emission residuals (blue shade) and the AGN \ion{Fe}{2} line
(yellow shade) are masked out before fitting.}
\label{fig:spectra}
\end{figure}

In Figure~\ref{fig:spectra} we present the normalized spectra of the 3 AGNs (black line) along 
with the best-fit template (red line). 
A good fit was obtained for NGC 3227 and Akn 120, while the $\sigstar$ of 3C 390.3
was not reliably determined since the fits was relatively poor. Thus, we discard the H-band $\sigstar$ 
measurement of 3C 390.3 and use the previous optical measurement from the literature in the further
analysis although our best fit H-band $\sigstar$ measurement is in close agreement with 
the optical measurement.
Compared to the previous $\sigstar$ measurements from optical spectra, our H-band measurements 
are slightly smaller (see Table 1).  Given the small sample size, however, we cannot conclude
whether there is a systematic difference between optical and near-IR measurements. 
Note that using a much larger sample of quiescent galaxies, we found that velocity dispersions 
measured in the optical and near-IR are in good agreement in our companion study (K13).  

\subsection{Rotation and Aperture Effect for NGC 3227} \label{section:roap}


\begin{figure}
\centering
\includegraphics[width = 0.47 \textwidth]{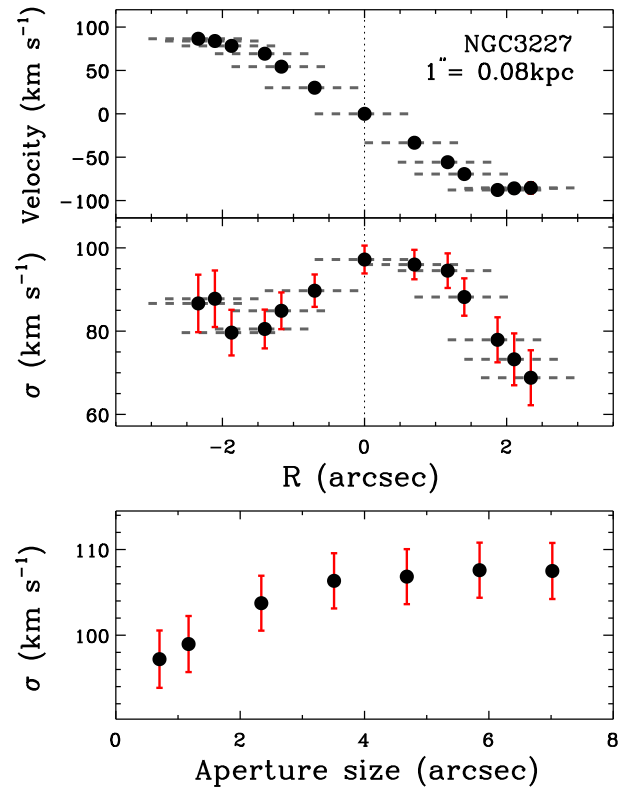}
\caption{Velocity (top panel) and velocity dispersion profiles (middle panel) of NGC~3227. 
The bottom panel illustrates the dependence of the measured $\sigstar$ on the extraction aperture 
size.}
\label{fig:curve}
\end{figure}

In the case of NGC 3227, we obtained spatially resolved spectra to measure velocity and velocity 
dispersion along the major axis of the host galaxy. The importance of obtaining spatially resolved 
spectra rather than single-aperture spectrum for a precise $\sigstar$ measurement has been
pointed out by previous studies \citep[e.g.,][]{Bennert:2011, Harris:2012}. In particular, K13 demonstrated 
that the velocity dispersion can be overestimated by up to $\sim$20\% if a single-aperture spectrum
is used without correcting for the rotational broadening. 

Fig.~\ref{fig:curve} presents the spatially resolved velocities and velocity dispersions measured 
along the major axis of NGC 3227. A clear rotation component is present with a maximum projected velocity
of $\sim$100\kms. Thus, the rotational broadening would lead to an increase of stellar velocity 
dispersion if a spectrum extracted from a large aperture is used.
To illustrate this, we measured $\sigstar$ from a series of spectra extracted from various aperture sizes
(Fig.~\ref{fig:curve} lower panel). As expected, the measured $\sigstar$ increases with
increasing aperture size due to the contribution of the rotation component.
The measured $\sigstar$  flattens after peaking around $\reff$, 
as the rotation curve flattens at $\sim$2" and the radial profile of the intrinsic $\sigstar$ decreases
outwards.
We corrected for the rotation effect by computing the luminosity-weighted $\sigstar$ within the effective 
radius,  
\begin{equation}
\sigma_{*}=\frac{\int_{-R_e}^{R_{e}} \sigstar(r)\, I(r)\,  {\rm d}r}{\int_{-R}^{R}  I(r)\,  {\rm d}r} \, \label{eq1}
\end{equation}
where $I(r)$ is the surface brightness profile of the galaxy measured from the spectral image
and $R_e$ is the effective radius.
We find the luminosity-weighted $\sigstar=92{\pm 6}$~km~s$^{-1}$, while the uncorrected value 
based on a single-aperture spectrum is $106{\pm 7}$~km~s$^{-1}$, i.e., we overestimate $\sigstar$ by 15\% if we do not correct for the rotation component.
Note that the rotation effect on $\sigstar$ measurements will vary for individual galaxies
since it depends on the inclination angle of the rotating disk to the line-of-sight and 
the maximum rotation velocities (see K13 for details).


\section{The \msigma\ Relation} \label{section:msig}

\begin{deluxetable}{lcc}
\tablecolumns{3}
\tablewidth{15pc}
\tablecaption{Updated $\mbh$ and $\sigstar$ measurements for quiescent galaxies}
\tablehead{
\colhead{Name}&\colhead{$M_{\rm BH}$}&\colhead{$\sigma_{\rm *}$}\\
\colhead{ }& \colhead{($10^8\, M_{\sun}$)} & \colhead{(km s$^{-1}$)}\\
\colhead{(1)}&\colhead{(2)}&\colhead{(3)}
}
\startdata
N221     &   $   0.026^{+0.005}_{-0.005}$  &    $  65\pm  2$ \\
N821     &   $   1.7^{  +0.7}_{  -0.7}$    &    $ 191\pm  8$ \\
N1023    &   $   0.40^{ +0.04}_{ -0.04}$   &    $ 205\pm  6$ \\
N2787    &   $   0.41^{ +0.04}_{ -0.05}$   &    $ 170\pm  4$ \\
N3031    &   $   0.80^{ +0.20}_{ -0.11}$   &    $ 165\pm  4$ \\
N3115    &   $   8.9^{  +5.1}_{  -2.7}$    &    $ 236\pm  9$ \\
N3227    &   $   0.15^{ +0.05}_{ -0.08}$   &    $  92\pm  6$ \\
N3245    &   $   2.1^{  +0.5}_{  -0.6}$    &    $ 192\pm  6$ \\
N3377    &   $   1.8^{  +0.9}_{  -0.9}$    &    $ 127\pm  3$ \\
N3379    &   $   4.2^{  +1.0}_{  -1.1}$    &    $ 205\pm  6$ \\
N3384    &   $   0.11^{ +0.05}_{ -0.05}$   &    $ 129\pm  4$ \\
N3607    &   $   1.4^{  +0.4}_{  -0.5}$    &    $ 198\pm  7$ \\
N3608    &   $   4.7^{  +1.0}_{  -1.0}$    &    $ 193\pm  5$ \\
N4258    &   $   0.367^{+0.001}_{-0.001}$  &    $ 109\pm  4$ \\
N4261    &   $   5.3^{  +1.1}_{  -1.1}$    &    $ 304\pm  8$ \\
N4291    &   $   9.8^{  +3.1}_{  -3.1}$    &    $ 245\pm  7$ \\
N4342    &   $   4.6^{  +2.6}_{  -1.5}$    &    $ 199\pm  8$ \\
N4374    &   $   9.2^{  +1.0}_{  -0.8}$    &    $ 292\pm  9$ \\
N4459    &   $   0.70^{ +0.13}_{ -0.14}$   &    $ 156\pm  7$ \\
N4473    &   $   0.89^{ +0.45}_{ -0.44}$   &    $ 173\pm  5$ \\
N4486    &   $  62^{    +3}_{    -4}$      &    $ 327\pm 11$ \\
N4564    &   $   0.88^{ +0.24}_{ -0.24}$   &    $ 166\pm  6$ \\
N4596    &   $   0.84^{ +0.36}_{ -0.25}$   &    $ 136\pm  5$ \\
N4649    &   $  47^{    +11}_{   -10}$     &    $ 311\pm 12$ \\
N4697    &   $   2.0^{  +0.2}_{  -0.2}$    &    $ 153\pm  4$ \\
N5845    &   $   4.9^{  +1.5}_{  -1.6}$    &    $ 223\pm  5$ \\
N6251    &   $   6.0^{  +2.0}_{  -2.0}$    &    $ 296\pm 13$ \\
N7052    &   $   4.0^{  +2.8}_{  -1.6}$    &    $ 334\pm 15$ 
\enddata
\label{tab:quiescentgal}
\tablecomments{Col. (1): galaxy name. Col. (2): $\mbh$ taken from \citet[][see references therein]{McConnell:2013}. (3): rotation-corrected stellar velocity dispersions taken from K13.}
\end{deluxetable}


\subsection{The \msigma\   Relation of Quiescent Galaxies} \label{section:msign}

Much effort has been invested to improve the \msigma\ relation since its discovery,
by increasing the sample size and the dynamical range \citep[e.g.][]{Gultekin:2009,Nowak:2010,McConnell:2012}, and by improving the previous $\mbh$ measurements \citep[e.g.][]{vandenBosch:2010,Shen:2010,Schulze:2011,Gebhardt:2011}. 
In contrast, $\sigstar$ measurements for many objects are still often based on rather 
inhomogeneous literature data. To overcome this limitation, we homogeneously measured
$\sigstar$ based on the uniformly taken H-band spectra for a sample of 28 quiescent galaxies,
as presented by K13, which is almost half the size of the most recent compilation of
$\mbh$ measurements \citep{McConnell:2013}. In particular, these $\sigstar$ measurements were 
corrected for the effect of rotation based on the spatially resolved measurements
using Equation 2, as similarly performed for NGC 3227 in Section 3.1.

Incorporating these new $\sigstar$ measurements from our companion work (K13), 
we here update the \msigma\ relationship for quiescent galaxies. 
The sample is based on the catalog of 72 galaxies presented by \citet{McConnell:2013},
of which we updated $\sigstar$ for 28 objects, taken from K13.
In Table 2, we list $\mbh$ and $\sigstar$ for the updated galaxies.
To fit the \msigma\ relationship we assumed the common single-index power law as expressed in Eq. 1,
and used two different fitting methods: the modified FITEXY method, accounting for intrinsic scatter 
in the fit \citep{Tremaine:2002}, and the maximum likelihood method \citep[][see \citet{Park:2012b} 
for more details on the comparison of these fitting techniques]{Gultekin:2009,Schulze:2011}. 
As we found consistent results between both methods, we only provide the results from the FITEXY method. 
For this method, we minimize 
\begin{equation}
 \chi^2 = \sum_{i=1}^N \frac{ \left( \mu_i -\alpha -\beta s_i \right)^2}{\sigma_{\mu,i}^2 + 
 \beta^2 \sigma_{s,i}^2 + \epsilon_0^2}   \  ,
\end{equation}
with $\mu=\log (\mbh/M_{\odot})$, $s=\log(\sigstar/ 200\,\mathrm{km\,s}^{-1})$,  
measurement uncertainties $\sigma_{\mu}$ and $\sigma_{s}$ in both variables, and 
intrinsic scatter $\epsilon_0$.
We obtain the best-fit result for the full sample and for completeness, we also fit 
the \msigma\ relation for each subsample, e.g., classical bulges, pseudo-bulges, core galaxies, 
and power-law galaxies, using classifications from the literature \citep{Kormendy:2011,Greene:2010,Gultekin:2009,McConnell:2013}. The best fit results are listed in Table 3,
showing that the slopes of the \msigma\ relation for these subsample 
are more or less similar except pseudo-bulge galaxies. The much shallower slope of the \msigma\ 
relation of pseudo-bulge galaxies is qualitatively consistent with the previous claims 
that pseudo-bulge galaxies do not follow the same \msigma\ relation as classical bulges 
\citep{Hu:2008,Graham:2008,Gadotti:2009,Greene:2010,Kormendy:2011}.

For the full sample, the slope ($\beta=5.31 \pm 0.33$) is slightly lower than that of 
\citet{McConnell:2013} but consistent within the uncertainties (see Figure 3). The difference is partly 
due to the method of calculating effective velocity dispersions. 
While we determined the effective velocity dispersion by calculating luminosity-weighted
velocity dispersion using Eq. 2, \citet[][see also \citet{Gultekin:2009}]{McConnell:2013}
added rotation velocity to velocity dispersion in quadrature
in calculating effective velocity dispersion within the effective radius, as
\begin{equation}
\sigma_{*}^2=\frac{\int_{-R_e}^{R_{e}} ( \sigstar(r)^2 + V(r)^2 ) \, I(r)\,  {\rm d}r}{\int_{-R_e}^{R_e}  I(r)\,  {\rm d}r} \, 
\end{equation}
where $I(r)$ is the surface brightness of the galaxy and $R_e$ is the effective radius.
Nevertheless, the change of the final adopted $\sigstar$ due to the
inclusion or exclusion of rotation velocity is not significant for early-type galaxies
because rotation velocity is relatively small and the central velocity dispersion is 
dominant in the luminosity-weight.
In contrast, the rotation effect is potentially important
for late-type galaxies, whose rotation velocity is comparable to velocity dispersion.
Note that we combined the rotation-corrected velocity dispersion for 28 galaxies
based on our previous work (K13) with the rotation-included velocity dispersion for 
the other 44 galaxies, for which rotation-corrected velocity dispersions are not available from 
\citet{McConnell:2013}.
To quantify the effect of rotation, we used rotation-included $\sigstar$ using Eq. 4
for the full sample, and fit the \msigma\ relation for various subsamples as listed in
Table 3. As the effective velocity dispersion increases due to the inclusion 
of rotation velocity, the slope of the \msigma\ relation slightly steepens for each subsample.

We find that late-type galaxies have a slightly shallower slope than early-type galaxies,
while \citet{McConnell:2013} claimed a consistent slope of the \msigma\ relation between 
early-type and late-type subsamples. As the effect of rotation is potentially much stronger 
in late-type galaxies than in early-type galaxies, it should be further tested 
whether early- and late-type galaxies have a consistent slope of the \msigma\ relation,
using a larger sample of late-type galaxies with/without rotation correction.
Note that rotation-corrected $\sigstar$ are only available for 3 late-type galaxies in the sample.
Applying such a correction to the remaining galaxies will mainly reduce $\sigstar$ 
for late-type and low $\sigstar$ galaxies, thus presumably flattening the slope of the \msigma\        
relationship. 

For understanding of the origin of the \msigma\ relation, 
the slope of the relation provides insight in the physical feedback process. 
For example, theoretical models propose a slope of $4$ for feedback by momentum-driven winds \citep{Fabian:1999,King:2003,Murray:2005}, while energy-driven winds lead to a slope of $5$ \citep{Silk:1998}. 
For the full sample, we find a slope $\sim5$, supporting  energy-driven winds. 
However, note that the rotation effect has not been corrected for 
most of low-mass and late-type galaxies. Thus, it is possible that the slope 
may be overestimated if $\sigstar$ of these late-type galaxies at low-mass 
scale were overestimated due to the rotation effect.
We caution that uncertainties in the slope are still large and selection effects will probably lead to a steeper slope in the \msigma\        relation \citep{Gultekin:2009,Schulze:2011b,Morabito:2012}.
We also note that the slope can significantly change depending on the fitting direction (forward vs. inverse regression) 
\citep[see Section 4.4 for more discussion;][]{Schulze:2011b,Graham:2011,Park:2012b}.


\begin{figure}
\centering
\includegraphics[width = 0.49 \textwidth]{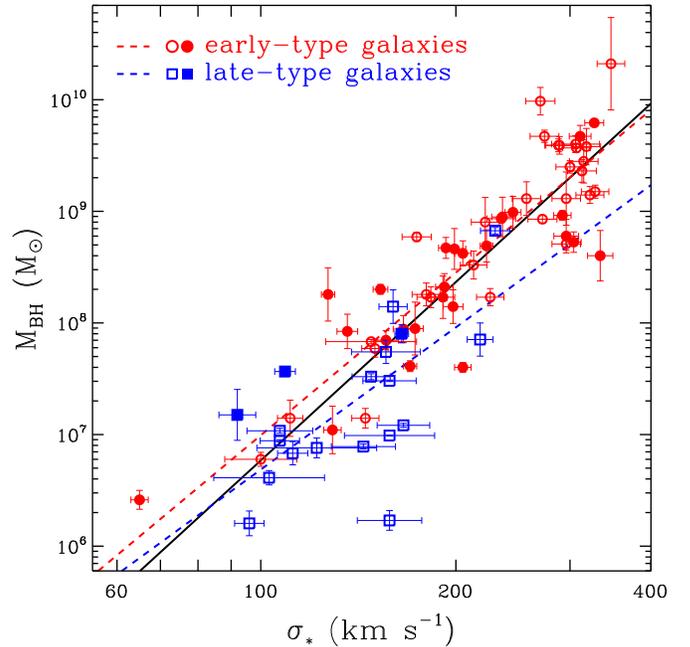}
\caption{\msigma\ relation of quiescent galaxies for the full sample (black line),
early-type galaxies (red circles, red dashed line), and late-type galaxies (blue squares, 
blue dashed line). The rotation-corrected H-band $\sigstar$ measurements are denoted with filled symbols
while the rotation-included $\sigstar$ values from the literature are marked with open symbols. 
}
\label{fig:msigma_quiescent}
\end{figure}

\begin{deluxetable*}{lccccccc}
\tablecolumns{8}
\tablewidth{0pc}
\tablecaption{Fits to the \msigma\ relationship}
\tablehead{
\colhead{Sample}&\colhead{$N_{\rm gal}$} & \multicolumn{3}{c}{rotation-corrected $\sigstar$} & \multicolumn{3}{c}{rotation-included $\sigstar$} \\
\colhead{ }&\colhead{ }&\colhead{$\alpha$}&\colhead{$\beta$}&\colhead{$\epsilon_0$} &\colhead{$\alpha$}&\colhead{$\beta$}&\colhead{$\epsilon_0$} 
}
\startdata
full          & 72 & $8.37{\pm 0.05}$ &  $5.31{\pm 0.33}$ & $0.41{\pm 0.05}$ & $8.33{\pm 0.05}$ &  $5.48{\pm 0.32}$ & $0.39{\pm 0.05}$ \\
early-type    &  53 & $8.45{\pm 0.05}$ & $4.84{\pm 0.36}$ & $0.35{\pm 0.04}$ & $8.40{\pm 0.06}$ & $5.04{\pm 0.36}$ & $0.35{\pm 0.04}$ \\
late-type     & 19 & $7.96{\pm 0.26}$ & $4.23{\pm1.26}$ & $0.48{\pm0.10}$    & $8.03{\pm 0.26}$ & $4.75{\pm1.21}$ & $0.45{\pm0.11}$ \\
classical bulges  & 54 & $8.49{\pm 0.05}$ & $4.57{\pm 0.33}$ & $0.33{\pm 0.04}$  & $8.45{\pm 0.05}$ & $4.76{\pm 0.33}$ & $0.33{\pm 0.04}$ \\
pseudo-bulges & 17 & $7.66{\pm 0.25}$ & $3.28{\pm 1.11}$ &  $0.35{\pm0.12}$ & $7.68{\pm 0.29}$ & $3.56{\pm 1.64}$ &  $0.33{\pm0.11}$ \\
core          & 28 & $8.55{\pm 0.09}$ & $4.67{\pm 0.69}$ &  $0.34{\pm0.06}$ & $8.55{\pm 0.10}$ & $4.66{\pm 0.74}$ &  $0.35{\pm0.06}$ \\
power-law     & 18 & $8.30{\pm 0.09}$ & $4.07{\pm 0.66}$ &  $0.37{\pm0.07}$ & $8.22{\pm 0.08}$ & $4.31{\pm 0.61}$ &  $0.34{\pm0.07}$ \\
barred        & 11 & $7.75{\pm 0.14}$ & $2.21{\pm 0.82}$ &  $0.32{\pm0.08}$ & $7.76{\pm 0.16}$ & $2.40{\pm 0.88}$ &  $0.31{\pm0.08}$ \\
non-barred    & 61 & $8.39{\pm 0.05}$ & $5.44{\pm 0.38}$ &  $0.39{\pm0.06}$ & $8.35{\pm 0.05}$ & $5.59{\pm 0.37}$ &  $0.37{\pm0.05}$ \\
\noalign{\smallskip} \hline \noalign{\smallskip}
reverberation-mapped AGN & 25 & $7.31{\pm 0.15}$ & $3.46{\pm 0.61}$ & $0.41{\pm 0.05}$  & & & \\
quiescent + AGN & 97 & $8.36{\pm 0.05}$ & $4.93{\pm 0.28}$ & $0.43{\pm 0.04}$ & & & 
\enddata
\tablecomments{Rotation-corrected velocity dispersions are available only for 28 objects in the
quescent galaxy sample from K13 (see Table 2) and for 1 object (NGC 3227) in the active galaxy sample
from this study.}
\label{tab:fits}
\end{deluxetable*}

\subsection{The $\mbh$-$\sigstar$ Relation of Active Galaxies}  \label{section:msiga}

We present the \msigma\ relation for reverberation-mapped AGNs using the sample compiled by \citet{Woo:2010} with 
the following updates: (1) improved $\sigstar$ measurements for 2 AGN (NGC~3227 and Akn~120), as presented in 
Section~\ref{section:anal}, (2) improved virial products for 9 AGNs from the Lick AGN Monitoring Project \citep{Park:2012a}, (3) addition of Mrk~50 \citep{Barth:2011}. 
It would be desirable to separate the AGN sample into subsamples, i.e., early/late-type and classical/pseudo bulges, 
as we did for the quiescent galaxy sample. However, obtaining a reliable classification for the host galaxies of broad line AGNs is hampered by the presence of a bright AGN \citep[e.g.,][]{Bentz:2009}, hence not available for the sample. 
Thus, we restrict our analysis to the full AGN sample although we note that our AGN sample is biased towards 
late-type galaxies and a significant fraction is expected to contain pseudo-bulges. 

Assuming the same form for the \msigma\ relation as for quiescent galaxies, we fit the relation 
for the AGN sample. Using the virial product in Table 4, we obtain $\alpha= 7.31 \pm 0.15$,  
$\beta=3.46 \pm 0.61$ and an intrinsic scatter of $\epsilon_0=0.41 \pm 0.05$. The slope is flatter than 
that of quiescent galaxies, but fully consistent with previous results on the \msigma\ relation of active galaxies. 
For example, \citet{Woo:2010} reported a slope of $\beta=3.55 \pm 0.6$ for their reverberation-mapped AGN sample. 
In the case of type-1 AGNs without reverberation mapping data, a similar slope, ranging from 3.3 to 3.7 has been 
reported by various studies \citep{Greene:2006, Shen:2008, Bennert:2011, Xiao:2011}, based on single-epoch $\mbh$ estimates and $\sigstar$ measurements. Thus, the apparent flattening of the \msigma\ relation for AGN samples is well established. 
In Section 4.4, we will address whether this is evidence for an intrinsic difference between active and inactive BHs and their galaxies, or it can be understood by differences in the sample selection.


\begin{deluxetable}{lcccc}
\tablecolumns{5}
\tablewidth{0pc}
\tablecaption{Virial Products and $\sigstar$ for reverberation AGN.}
\tablehead{
\colhead{Name}&\colhead{VP}&\colhead{Ref.}&\colhead{$\sigma_{\rm *}$}&\colhead{Ref.}\\
\colhead{}&\colhead{$\sigma_{\rm line}^{2}$R$_{BLR}$/G}&\colhead{}&\colhead{}&\colhead{}\\
\colhead{ }& \colhead{(10$^{6}$$M_{\sun}$)} & \colhead{} & \colhead{(km s$^{-1}$)} & \colhead{}\\
\colhead{(1)}&\colhead{(2)}&\colhead{(3)}&\colhead{(4)}&\colhead{(5)}
}
\startdata
Akn 120                & $27.2{\pm 3.5}$                         &2& $192{\pm 8}$ &This work\\
Arp 151                 & $1.31_{\rm -0.23}^{+0.18}$     &3& $118{\pm 4}$ &7\\
Mrk 50                   & $6.2{\pm 0.9}$                            &6& $109{\pm 14}$ & 6\\
Mrk 79                   & $9.52{\pm 2.61}$                       &2& $130{\pm 12}$ &9\\
Mrk 110                 & $4.57{\pm 1.1}$                         &2& $91{\pm 7}$ &10\\
Mrk 202                 & $0.55_{\rm -0.22}^{+0.32}$     &3& $78{\pm 3}$ &7\\
Mrk 279                 & $6.35{\pm 1.67}$                      &2& $197{\pm 12}$ &9\\
Mrk 590                 & $8.64{\pm 1.34}$                      &2& $189{\pm 6}$ &9\\
Mrk 817                 & $11.3_{\rm -2.8}^{+2.7}$         &1& $120{\pm 15}$ &9\\
Mrk 1310              & $0.61{\pm 0.20}$                       &3& $84{\pm 5}$ &7\\
NGC 3227            & $1.39_{\rm -0.31}^{+0.29}$     &1& $92{\pm 6}$ &This work\\
NGC 3516            & $5.76_{\rm -0.76}^{+0.51}$     &1& $181{\pm 5}$ &9\\
NGC 3783            & $5.42{\pm 0.99}$                       &2& $95{\pm 10}$ &11\\
NGC 4051            & $0.31_{\rm -0.09}^{+0.10}$     &1& $89{\pm 3}$ &9\\
NGC 4151            & $8.31_{\rm -0.85}^{+1.04}$     &4& $97{\pm 3}$ &9\\
NGC 4253            & $0.35_{\rm -0.14}^{+0.15}$     &3& $93{\pm 32}$ &7\\
NGC 4593            & $1.78{\pm 0.38}$                       &5& $135{\pm 6}$ &9\\
NGC 4748            & $0.68_{\rm -0.30}^{+0.24}$     &3& $105{\pm 13}$ &7\\
NGC 5548            & $12.41_{\rm -4.21}^{3.06}$     &3& $195{\pm 13}$ &7\\
NGC 6814            & $3.73_{\rm -1.11}^{+1.10}$     &3& $95{\pm 3}$ &7\\
NGC 7469            & $2.21{\pm 0.25}$                       &2& $131{\pm 5}$ &9\\
SBS 1116+583A & $1.08_{\rm -0.49}^{+0.52}$     &3& $92{\pm 4}$ &7\\
PG 1426+015      & $236{\pm 70}$                           &2& $217{\pm 15}$ &12\\
3C 120                  & $10.1_{\rm -4.1}^{+5.7}$          &2& $162{\pm 20}$ &8\\
3C 390.3              & $52.2{\pm 11.7}$                       &2& $273{\pm 16}$ &9 
\enddata
\label{table4}
\tablecomments{Col. (1): galaxy name. Col. (2): virial product ($\mbh = f\times$VP). 
Col. (3): reference for $\mbh$ Col. (6): stellar velocity dispersion Col. (7): reference for $\sigstar$\\
References. --- (1) \citet{Denney:2010}; (2) \citet{Peterson:2004}; (3) \citet{Park:2012a}; (4) \citet{Bentz:2006}; (5) \citet{Denney:2006}; (6) \citet{Barth:2011}; (7) \citet{Woo:2010}; (8) \citet{Nelson:1995}; (9) \citet{Nelson:2004}; (10) \citet{Ferrarese:2001}; (11) \citet{Onken:2004}; (12) \citet{Watson:2008}} 
\end{deluxetable}


\begin{figure}
\centering
\includegraphics[width = 0.49 \textwidth]{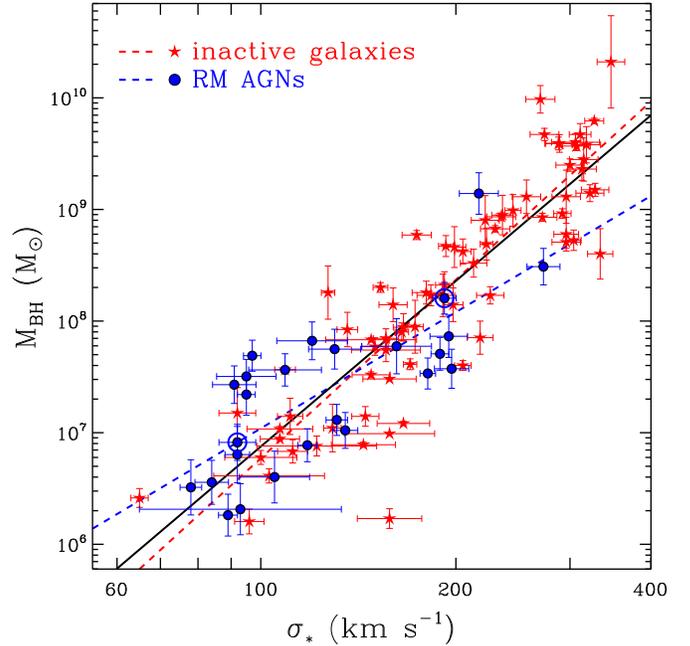}
\caption{\msigma\ relation for active (blue circles) and quiescent galaxies 
(red stars). The 2 AGNs with updated $\sigstar$ are marked 
with big blue circles. The red dashed line represents the best fit for quiescent galaxies, as in Fig.~\ref{fig:msigma_quiescent}, the blue dashed line shows the relation for the reverberation-mapped AGNs 
(assuming $f=5.9^{+2.1}_{-1.5}$ based on the joint fit), and the black solid line shows the 
best fit to the combined quiescent and active galaxy samples}
\label{fig:msigma_agn}
\end{figure}

\subsection{The Virial Factor}  \label{section:ffactor}
For type~1 AGNs reverberation mapping provides an alternative method to measure $\mbh$. 
However, apart from a few recent exceptions \citep{Brewer:2011,Pancoast:2012}, only 
the virial product (VP) can be obtained which deviates from the actual $\mbh$ by an 
unknown normalization factor $f$ as expressed
in Sec. 1. This $f$ factor is either determined by assumptions about the geometry of the BLR \citep[e.g.][]{Netzer:1990}, or by scaling of the reverberation mapping sample to the \msigma\        
relation of quiescent galaxies \citep{Onken:2004,Woo:2010}. A fully spherical BLR corresponds to a virial factor $f=3$  \citep[e.g.][]{Kaspi:2000}, while if the BLR is modeled by a rotating disk, the virial factor is viewing angle dependent \citep[e.g.][]{Collin:2006}. Scaling the reverberation mapping sample to the quiescent galaxy \msigma\        relation is an independent approach to determine the average virial factor $f$. The underlying assumption of this approach is that AGNs obey the same relation as quiescent galaxies. 

Based on a sample of 16 reverberation mapping AGN, \citet{Onken:2004} reported a virial factor of $f=5.5$, adopting the relations from  \citet{Tremaine:2002} and  \citet{Ferrarese:2002}. \citet{Woo:2010} obtained $f=5.2$ and $f=5.1$, calibrating an enlarged sample of 24 AGN to the relations of \citet{Gultekin:2009} and \citet{Ferrarese:2005}, respectively. 
\citet{Graham:2011} reported a smaller value of $f=3.8$ by calibrating the AGN sample to the $\mbh$-$\sigstar$ relation for quiescent galaxies presented in their paper. 
Recently, \citet{Park:2012b} investigated the origin of this discrepancy in the determination of $f$ in the last two studies. They concluded that the difference is mainly due to the sample selection in both studies. Indeed, the $\mbh$-$\sigstar$ relation reported by \citet{Graham:2011} is offset by $0.2$~dex compared to our results in Section~\ref{section:msign}. Using the $\mbh$-$\sigstar$ from \citet{McConnell:2011}, \citet{Park:2012b} obtained $f=5.1$, in excellent agreement with the results of \citet{Woo:2010}.

We here provide a new determination of the virial factor, based on the updated reverberation mapping AGN sample presented in section~\ref{section:msiga}, calibrated to the quiescent galaxy sample presented in section~\ref{section:msign}. We fitted the AGN sample with the FITEXY and maximum likelihood method, but fixed the slope to $\beta=5.31$, found for the quiescent galaxy sample. As we were using only virial products instead of $\mbh$ for the AGN sample, the virial factor $f$ is given by $\log f = \alpha_{\rm quiescent} - \alpha_{\rm AGN}$. Using the FITEXY method, we obtain a virial factor $\log f=0.71 \pm 0.11$, corresponding to  $f=5.1_{-1.1}^{+1.5}$, and intrinsic scatter $\epsilon_0=0.49 \pm 0.05$, consistent with most previous results. 

As we have shown above, the slope of the AGN sample is not consistent with the slope found for the quiescent galaxy sample. This may potentially lead to a bias on the determined virial factor. If we hold to the assumption that quiescent and active galaxies follow the same $\mbh$-$\sigstar$ relation, the most consistent approach is to fit the quiescent galaxy sample and the AGN sample together, determining the $\mbh$-$\sigstar$ relation and the virial factor jointly. 
This can be easily achieved by the maximum likelihood method, where we minimize the likelihood  $S=-2\ln \mathcal{L}$. In our case, the likelihood function $S$ is then the sum of the likelihood functions for the quiescent and active galaxies,
\begin{eqnarray}
 S  & = & \sum_{i=1}^N \left[ \frac{ \left( \mu_i -\alpha -\beta s_i \right)^2}{\epsilon_{\mathrm{tot},i}^2}  + 2 \ln \epsilon_{\mathrm{tot},i} \right]  + \\ 
 & &  \sum_{j=1}^M \left[ \frac{ \left( \mu_{{\rm VP},j}  +\log f-\alpha -\beta s_j \right)^2}{\epsilon_{\mathrm{tot},j}^2+\epsilon_f^2}  + 2 \ln \sqrt{\epsilon_{\mathrm{tot},j}^2+\epsilon_f^2} \right] \ , \nonumber
\end{eqnarray}
where $\mu_{{\rm VP}}=\log {\rm VP}$, $\epsilon_{\mathrm{tot}}^2=\sigma_{\mu}^2+\beta^2\sigma_{s}^2+\epsilon_0^2$, $N$ is the number of  galaxies with dynamical mass measurements and $M$ is the number of AGN with reverberation mapping masses. We simultaneously fit for the zero point $\alpha$, slope $\beta$ and intrinsic scatter $\epsilon_0$ of the \msigma\        relation and for the mean virial factor $\log f$ and its intrinsic scatter  $\epsilon_f$.  Our best fit results are $\alpha=8.36 \pm 0.05$,  $\beta=4.78 \pm 0.26$,  $\epsilon_0=0.42 \pm 0.04$ and $\log f=0.79 \pm 0.13$, consistent with our previous results. The data do not show evidence for an additional intrinsic scatter in the $f$ value. This may suggest an intrinsic small spread in the virial factor, for example due to small variation in the viewing angle onto a non-spherical BLR. Another possible explanation for the apparent lack of scatter in $f$ 
is that measurement uncertainties are under- or overestimated in the quiescent and/or active galaxy 
samples.

Assuming zero intrinsic scatter in the $f$ value, we can also use a modified version of the FITEXY method to obtain the $\mbh$-$\sigstar$ relation and the $f$ factor jointly. Here, we minimize
\begin{equation}
 \chi^2 = \sum_{i=1}^N \frac{ \left( \mu_i -\alpha -\beta s_i \right)^2}{\sigma_{\mu,i}^2 + 
 \beta^2 \sigma_{s,i}^2 + \epsilon_0^2}  + \sum_{j=1}^M  \frac{ \left( \mu_{{\rm VP},j}  +\log f-\alpha -\beta s_j \right)^2}{\sigma_{\mu,j}^2 + 
 \beta^2 \sigma_{s,j}^2 + \epsilon_0^2}\  ,
\end{equation}
where we change $\epsilon_0$ such that we obtain a reduced $\chi^2$ of unity. The best fit using this modified FITEXY is $\alpha=8.36 \pm 0.05$,  $\beta=4.93 \pm 0.28$,  $\epsilon_0=0.43 \pm 0.04$ and $\log f=0.77 \pm 0.13$, fully consistent with the results above. This corresponds to a virial factor $f=5.9_{-1.5}^{+2.1}$.

\subsection{Are the $\mbh$-$\sigstar$ relations of quiescent and active galaxies different?}
From a physical perspective it is not clear whether active galaxies should follow exactly the 
same \msigma\ relation as inactive galaxies, as they are in a special evolutionary phase of ongoing BH growth 
although the BH growth rate may not be very high. 
To first order, the overlap between the two samples, seen in Fig.~\ref{fig:msigma_agn}, suggests 
that active and inactive galaxies do obey the same relation. As discussed above, however,
the zero-point of the relation is ensured by design. In contrast, the measured slopes of the two relations 
appear to be mildly inconsistent with each other, implying that quiescent and 
active galaxies have different $\mbh$-$\sigstar$ relations.

Note that both samples rely on very different methods of estimating $\mbh$, and it is not well tested whether 
both methods give the same mass measurements. Currently only for two objects (NGC~3227 and NGC~4151),
$\mbh$ measurements are available from both dynamical method {\it and} reverberation mapping technique,
which are in reasonable agreement \citep{Davies:2006,Denney:2010, Onken:2007,Hicks:2008,Bentz:2006}.
However, more such cases are required to draw firm conclusions whether dynamical method and reverberation mapping 
technique provide consistent results. 

How can we then understand the apparent difference in the slope of the  $\mbh$-$\sigstar$ relation
between active and quiescent galaxy samples? It may imply a real physical difference, for example caused by a different evolutionary stage in the BH growth phase, where active BHs in less massive galaxies are still in the process of 
growing towards the quiescent $\mbh$-$\sigstar$ relation.
Alternatively,  the apparent difference could be simply due to the sample selection. In fact, galaxies studied with dynamical methods and those studied via reverberation mapping obey different selection criteria. The main observational limitation for dynamical methods is the requirement to approximately spatially resolve the BH's sphere of influence, given by $R_{\rm inf}=G\mbh \sigma_\ast^{-2}$. This naturally limits the applicability of this method to nearby galaxies, but also excludes a specific subset within this local volume. BHs with smaller
sphere of influence 
are removed from the sample, either implicitly by the target selection or through the ability to detect the BH \citep{Batcheldor:2010,Schulze:2011b}. 

Contrary, reverberation mapping does not depend on spatial resolution. However, it requires the presence of broad AGN emission lines and sufficient AGN variability to first classify the galaxy as harbouring an AGN and secondly measure a secure time lag. This introduces several selection effects to the AGN $\mbh$-$\sigstar$ sample,
including an active fraction bias and a luminosity bias as discussed by \citet{Schulze:2011b}. 
An active fraction bias is introduced if lower mass BHs have a higher probability to be in an active state. 
In contrast, low luminosity AGN with weak broad lines \citep[e.g.][]{Ho:1997, Greene:2007,Dong:2012}, will not be included in the reverberation mapping sample, causing a mild luminosity bias that can flatten the slope of 
the $\mbh$-$\sigstar$ relation. The expected magnitude of the combined effect on the reverberation-mapped AGN
sample is however small \citep[see for details][]{Schulze:2011b}.

The variability criterion can also lead to a selection effect since AGN luminosity variability is a key requirement 
for a successful reverberation mapping measurements. As there is an anti-correlation between the variability amplitude
and AGN luminosity \citep{Cristiani:1997,VandenBerk:2004}, brighter AGN will be preferentially excluded 
from reverberation mapping campaigns, introducing a bias against higher mass BHs. 

Probably the dominating selection effect for the $\mbh$-$\sigstar$ relation of AGN samples is the observational 
ability to measure a reliable $\sigstar$ of host galaxies.
This is particularly challenging for bright AGNs, where the host galaxy spectrum is swamped by the AGN continuum. 
Thus, only few reliable $\sigstar$ measurements exist for $\mbh>10^8\,M_\sun$, leading to a selection
bias in the AGN sample. Indeed, the current reverberation mapped AGN sample is skewed towards lower mass BHs, 
with an apparent lack of $\sigstar$ measurements for high mass BHs, limiting 
the dynamical range of the AGN sample, compared to the quiescent galaxy sample.
The combination of these effects, in particular the preferential exclusion of high mass BHs, 
will naturally lead to a flattening of the slope of the observed $\mbh$-$\sigstar$ relation.

We performed a simple Monte Carlo experiment to test the hypothesis: 
if the $\mbh$ distribution of the quiescent galaxies is limited
to the $\mbh$ distribution of the AGN sample, would the quiescent galaxy sample show a similar 
flattened slope? 
We constructed a large number of Monte Carlo samples, for each using the $\mbh$ of the 25 reverberation 
mapped AGNs. Then for each object we assign a $\sigstar$ value of a galaxy,
which is randomly chosen from the quiescent galaxy sample within a small $\mbh$ bin ($\sim 0.3$ dex) 
around the AGN's $\mbh$.  
Then, we fitted the $\mbh$-$\sigstar$ relations of the Monte Carlo samples. 
Based on 1000 Monte Carlo realizations, we find a slope of $\beta=3.81 \pm 0.48$, 
consistent with that of the AGN sample, implying that the limited $\mbh$ distribution causes
the flattening of the \msigma\ relation. If we revert the same experiment by 
sampling $\sigstar$ from the AGN sample and then assign $\mbh$ from a galaxy 
in the quiescent galaxy sample, 
we find a slope of $\beta=5.09 \pm 0.50$, consistent with the quiescent galaxy sample. 

We conclude that the AGN sample has a  $\sigstar$ distribution consistent with the quiescent galaxy $\sigstar$ 
distribution, however, it differs in its $\mbh$ distribution.  
Given the same  $\mbh$ distribution as the AGN sample, the quiescent galaxy sample would
follow the same $\mbh$-$\sigstar$ relation as the AGN sample. 
Thus, the difference between their apparent relations can be mainly attributed to 
sample selection effects. At a given $\sigstar$, dynamical methods and reverberation mapping of AGNs probe different regimes in $\mbh$. Some BHs will not be observed with dynamical methods, because their sphere of influence is not resolved. On the other hand, 
some active galaxies will not enter the sample, because of the various selection effects discussed above.

This is also consistent with the results for the fit of the inverse $\mbh$-$\sigstar$ relation for active and quiescent galaxies, presented by \citet{Park:2012b}. They fitted the $\mbh$-$\sigstar$ for both samples, first with the forward regression ($p(\mbh | \sigstar)$) and second with an inverse regression ($p(\sigstar | \mbh)$). 
The inverse regression has also been used by \citet{Graham:2011} to derive the $\mbh$-$\sigstar$ relation of quiescent
galaxies in order to avoid selection effects in the $\mbh$ distribution.
Indeed, the inverse fit is not affected by selection effects in the $\mbh$ distribution. 
It therefore can be a powerful tool to reconstruct the underlying relation, free of selection biases, 
in particular for active galaxies, as demonstrated by \citet{Schulze:2011b}. 
However, contrary to the forward regression, the inverse regression does not directly yield the intrinsic 
$\mbh$-$\sigstar$ relation, since it determines the conditional probability  $p(\sigstar | \mbh)$ of the bivariate distribution of $\mbh$ and $\sigstar$ along an orthogonal direction, independent of any additional selection effects \citep[][Schulze \& Wisotzki, in prep.]{Schulze:2011b}. Reconstructing the intrinsic relation from it requires the knowledge of 
the bulge distribution function and the intrinsic scatter in the  $\mbh$-$\sigstar$ relation. 
The exponential decrease of the bulge distribution function at high $\sigstar$ in combination with intrinsic scatter in the  $\mbh$-$\sigstar$ relation will lead to a deviation of the inverse relation from the intrinsic relation towards the high $\mbh$ end \citep{Schulze:2011b}. This upturn naturally causes a steeper slope when fitted by a single power law, consistent with observations \citep{Graham:2011,Park:2012b}.
Focusing on the strength of the inverse regression to avoid a selection bias on $\mbh$, we note that \citet{Park:2012b} 
reported similar slopes of the $\mbh$-$\sigstar$ relation for active and quiescent galaxies using the inverse regression,
while for the forward regression they found a difference in the slopes.
This supports our argument that the sample difference is in the range of $\mbh$ at a given $\sigstar$, while the range of $\sigstar$ at a given $\mbh$ in both samples is similar.



\section{Conclusions} \label{section:conclu}

To determine and compare the $\mbh$-$\sigstar$ relationship, we presented updated samples 
for quiescent and active galaxies, respectively, by combining our new stellar velocity 
dispersion measurements and previous measurements from the literature. 
While the quiescent galaxy sample is based on the compilation of \citet{McConnell:2013}, 
the main update is the addition of new, homogeneously 
measured stellar velocity dispersions for 28 galaxies, based on the spatially resolved H-band 
spectra obtained with the Triplespec at the Palomar 5m Hale telescope \citep{Kang:2013}.  
These $\sigstar$ measurements were corrected for the contribution of galaxy rotation through 
calculating luminosity-weighted velocity dispersion. The best-fit $\mbh$-$\sigstar$ relation of
quiescent galaxies shows a slope $\beta=5.31\pm0.33$ and an intrinsic scatter $\epsilon=0.41\pm0.05$. 

For the reverberation-mapped AGN sample, we improved $\sigstar$ measurements for 2 objects,
including NGC~3227, for which we also measured $\sigstar$ after correcting for
the rotation effect based on the spatially resolved measurements.
By updating $\sigstar$ for these 2 AGNs, we compiled a sample of 25 AGNs with reverberation $\mbh$ and 
reliable $\sigstar$ measurements \citep{Woo:2010}, in oder to compare with the quiescent galaxy
sample and to determine the virial factor $f$, which is required for the computation of virial $\mbh$.
First, we determine the virial factor $f$ by matching the AGN sample to our best fit  $\mbh$-$\sigstar$ 
relation of the quiescent galaxy sample. Second, we presented a new method to obtain the virial 
factor $f$, which is to fit simultaneously the $\mbh$-$\sigstar$ relation of quiescent and active
galaxies. Both methods provided consistent results, with a virial factor of $f=5.1_{-1.1}^{+1.5}$ and 
$f=5.9_{-1.5}^{+2.1}$, respectively.

When we determined the $\mbh$-$\sigstar$ relationship for the active galaxy sample alone, the slope of the 
$\mbh$-$\sigstar$ is shallower ($\beta=3.48\pm0.62$) than that of quiescent galaxies, 
in agreement with previous studies. However, due to the increase of the slope for the 
quiescent galaxies presented in this paper and other recent studies \citep{Graham:2011,Park:2012b,McConnell:2013}, 
the two slopes now are consistent with each other only on the $2\sigma$ level.
While the uncertainty in the slope is still large, this could be evidence for different  
$\mbh$-$\sigstar$ relations between quiescent and active galaxies. 
However, this apparent deviation is resolved when we consider selection effects, inherent in the 
observed samples. The reverberation-mapped AGN sample is slightly biased against low luminosity AGN 
with weak broad emission lines, hence, low mass BHs. However, more important is a bias against luminous 
AGN, harboring on average high mass BHs. This is due to a variability bias and a selection bias
due to $\sigstar$ measurement. The former is due to the anti-correlation between variability amplitude 
and AGN luminosity, while the latter is caused by the observational challenge to measure $\sigstar$ in the presence of 
a bright nuclear point source. This bias in the $\mbh$ distribution gives rise 
to the observed flattening in the slope. Accounting for the different $\mbh$ distributions 
determined by the dynamical method and by the reverberation mapping method, respectively, 
we find good agreement of the $\mbh$-$\sigstar$ relations between active and quiescent galaxies. 
This result assures the use of broad line AGNs as cosmological tools to probe cosmic 
evolution of the $\mbh$-$\sigstar$ relation out to high redshift \citep[e.g.,][]{Woo:2006,Woo:2008,Merloni:2010,Wang:2010,Bennert:2011}. At the same time, it emphasizes the need to account for sample selection effects 
in the design and interpretation of these studies.

\acknowledgements
We thank the anonymous referee for constructive suggestions.
This work was supported by the National Research Foundation of Korea (NRF) grant funded by the Korea government (MEST) (No. 2012-006087). J.H.W acknowledges the support by the Korea Astronomy and Space Science Institute (KASI) grant funded 
by the Korea government (MEST).

\clearpage

\end{document}